\documentclass[prl,aps,twocolumn,superscriptaddress,showpacs]{revtex4}
\usepackage{amsmath,amssymb,graphicx}
\usepackage{pxfonts}
\usepackage{graphicx}
\usepackage{color}
\usepackage{float}
\begin{document}

\date{\today}

\title{Fermi edge singularity in neutral electron-hole system}

\author{D.~J.~Choksy$^*$}
\affiliation{Department of Physics, University of California at San Diego, La Jolla, CA 92093, USA}
\author{E.~A.~Szwed$^*$}
\affiliation{Department of Physics, University of California at San Diego, La Jolla, CA 92093, USA}
\author{L.~V.~Butov} 
\affiliation{Department of Physics, University of California at San Diego, La Jolla, CA 92093, USA}
\author{K.~W.~Baldwin}
\affiliation{Princeton University, Princeton, New Jersey 08544, USA}
\author{L.~N.~Pfeiffer}
\affiliation{Princeton University, Princeton, New Jersey 08544, USA}

\begin{abstract}
\noindent
In neutral dense electron-hole (e-h) systems at low temperatures, theory predicts Cooper-pair-like excitons at the Fermi energy and a BCS-like exciton condensation. Optical excitation allows creating e-h systems with the densities controlled by the excitation power. However, the intense optical excitations required to achieve high densities cause substantial heating of the e-h system that prevents the realization of dense and cold e-h systems in conventional semiconductors. In this work, we study e-h systems created by optical excitation in separated electron and hole layers. The layer separation increases the e-h recombination time and, in turn, the density for a given optical excitation by orders of magnitude and, as a result, enables the realization of the dense and cold e-h system. We found a strong enhancement of photoluminescence intensity at the Fermi energy of the neutral dense ultracold e-h system that evidences the emergence of excitonic Fermi edge singularity due to the Cooper-pair-like excitons at the Fermi energy. 
\end{abstract}

\maketitle

The theory of an ultracold neutral electron-hole (e-h) system considers two density regimes. At low e-h densities, $n << 1/a_{\rm B}^D$ ($a_{\rm B}$ is the exciton Bohr radius, $D$ the dimensionality), electrons and holes bind to hydrogen-like pairs -- excitons, which form a Bose-Einstein condensate at low temperatures~\cite{Keldysh1968}. In dense electron-hole systems, $n \gtrsim 1/a_{\rm B}^D$, e-h plasma can be realized and, at low temperatures, the theory predicts Cooper-pair-like excitons at the Fermi energy and a BCS-like exciton condensation~\cite{Keldysh1965}.

\begin{figure}
\begin{center}
\includegraphics[width=5.5cm]{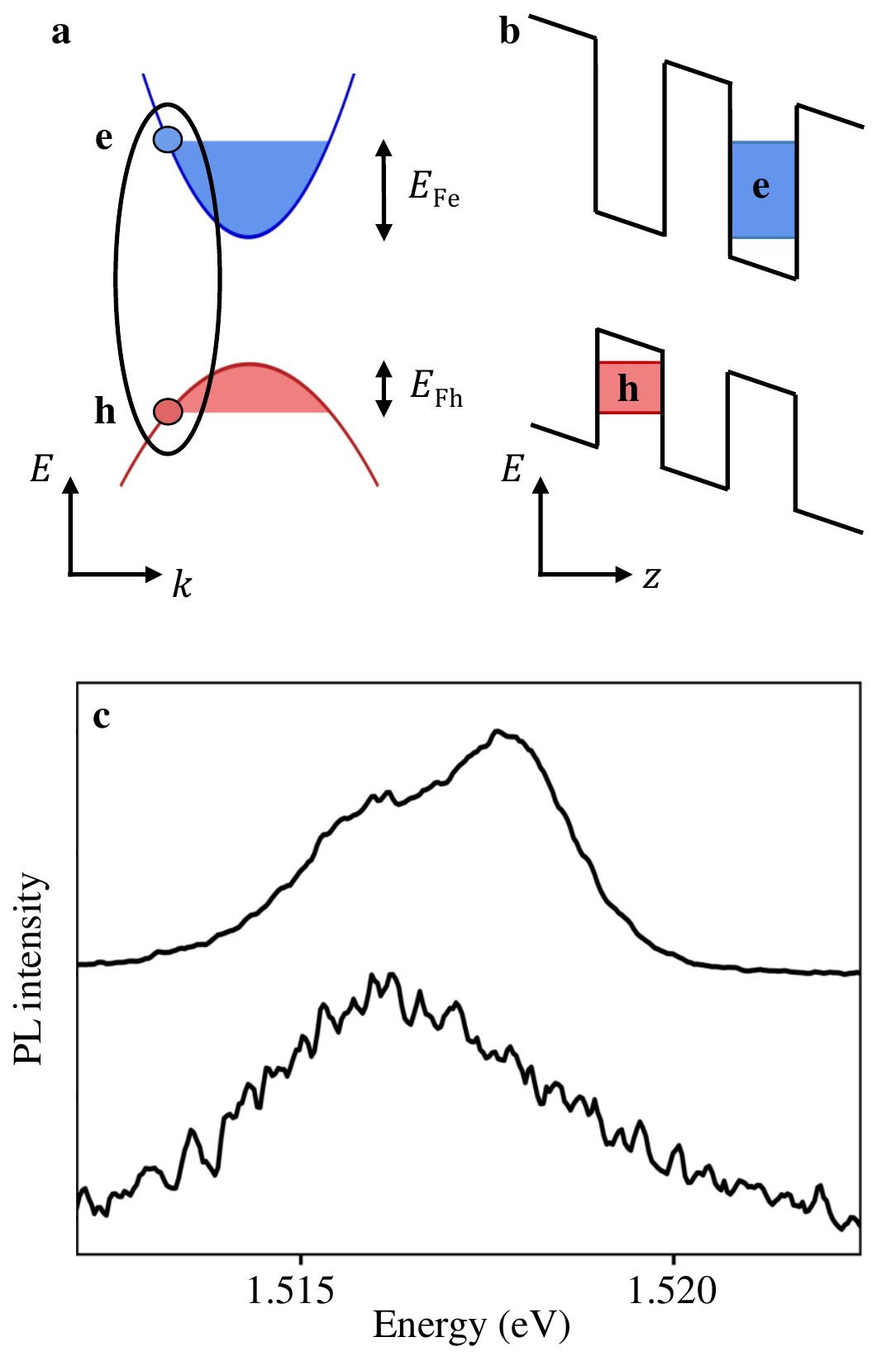}
\caption{(a) Diagram showing Cooper-pair-like excitons (oval) with electrons (e) and holes (h) at the Fermi energy in neutral dense e-h system. (b) Diagram of the CQW heterostructure. Electrons and holes in I-EHP are confined in separated layers. (c) I-EHP PL spectra at $T = 2$~K (top) and 20~K (bottom). $P_{\rm ex} = 24$~mW. The Fermi edge singularity is observed in cold I-EHP.
}
\end{center}
\label{fig:spectra}
\end{figure}

E-h systems can be created by optical excitation. The advantage of the optically created e-h systems is the ability to tune the density and, in turn, the parameter $na_{\rm B}^D$ within orders of magnitude by the excitation power $P_{\rm ex}$. In particular, this allows exploring both the BEC and BCS-like exciton condensates and the crossover between them. 2D e-h systems in layered semiconductor heterostructures offer an additional advantage of tailoring the system parameters by the layer design.

For a dense 2D e-h plasma with $n \gtrsim 1/a_{\rm B}^2$ at temperatures lower than the electron and hole Fermi energies and higher than the condensation temperature, the photoluminescence (PL) spectrum is step-like corresponding to the step-like 2D density of states, and the PL linewidth is approximately the sum of the electron and hole Fermi energies, $\Delta \sim E_{\rm Fe} + E_{\rm Fh}$~\cite{Butov1991}. In contrast, in the low-density regime, the exciton PL linewidth is determined by the homogeneous and inhomogeneous broadening and is significantly smaller than the e-h plasma PL linewidth in high-quality heterostructures~\cite{High2009}. The BCS-like condensation in the ultra-cold e-h plasma with Cooper-pair-like excitons at the Fermi energy should be accompanied by a PL intensity enhancement at the Fermi level of the step-like e-h plasma PL line, similar to the Fermi edge singularity in a Fermi-gas of electrons~\cite{Mahan1967, Skolnick1987}. The latter phenomenon was observed in PL spectra of a 2D electron gas for a weak optical excitation with the number of photoexcited electrons and holes much smaller than the electron gas density, that is in a system, which can be described as a single optically created hole in a Fermi sea of electrons. 

In contrast to the generation of "a single hole" in a dense electron gas~\cite{Skolnick1987}, the realization of a neutral dense e-h system by optical excitation requires the generation of a high number of electrons and holes that causes the problem of heating. Due to e-h recombination, the temperature of an optically created e-h system ($T_{\rm eh}$) exceeds the semiconductor lattice temperature and lowering $T_{\rm eh}$ below the condensation temperature, in particular at high e-h densities, is challenging. For instance, for neutral dense plasmas generated in single InGaAs/InP~\cite{Butov1991} or InGaAs/GaAs~\cite{Butov1992, Kappei2005} quantum wells in experiments at $T \sim 2$~K, the effective e-h temperature reached and exceeded $\sim 100$~K, well above both the lattice temperature and the temperature needed for the realization of BCS-like exciton condensation~\cite{Keldysh1965}. 

The other requirement for the realization of Cooper-pair-like excitons at the Fermi energy and BCS-like exciton condensation is matching of the electron and hole Fermi surfaces~\cite{Keldysh1965}. For equal electron and hole densities, the Fermi momenta of electrons and holes are equal (Fig.~1a), that is required for matching the Fermi surfaces. This matching of the electron and hole Fermi surfaces in neutral e-h system is different from the Fermi edge singularity for a hole in a Fermi sea of electrons where the suppression of hole kinetic energy, a flat hole band or hole localization, is required~\cite{Mahan1967, Skolnick1987}. 

To create cold e-h systems, we work with heterostructures with separated electron and hole layers (Fig.~1b). In these heterostructures, spatially indirect excitons (IXs), also known as interlayer excitons, are formed by electrons and holes confined in separated layers. The layer separation increases the e-h recombination time that allows cooling the optically generated e-h system to low temperatures~\cite{Butov2001}. The other advantage of the separated electron and hole layers is the overall enhancement of energy per e-h pair with density that is outlined below. This enhancement stabilizes the exciton state against the formation of e-h droplets~\cite{Yoshioka1990, Zhu1995, Lozovik1997, deLeon2001}, which otherwise may form the ground state~\cite{Keldysh1986}. Earlier studies of cold IXs concerned the low-density regime where IXs are hydrogen-like pairs. An overview of experimental studies of IX condensation in the low-density regime and phenomena in the IX condensate can be found in Ref.~\cite{Leonard2021}. 

In this work, we study ultracold neutral spatially indirect e‐h plasma (I-EHP) in separated electron and hole layers in a GaAs/AlGaAs coupled quantum well (CQW) heterostructure. The electrons and holes are confined in 15~nm GaAs QWs separated by 4~nm AlGaAs barrier. The long e‐h recombination lifetimes [$\tau \sim \mu$s, Fig.~S1a in Supplementary Information (SI)] due to the separation between the electron and hole layers allow for cooling the plasma to low temperatures. The creation of cold I-EHP is facilitated by separating the e-h plasma from the laser excitation in space and time: (i) The measurements are performed $\delta t = 300$~ns after the laser excitation pulse within $\tau_{\rm w} = 50$~ns window (Fig.~S1a). This delay $\delta t$ is sufficient for cooling the photoexcited e-h system to low temperatures close to the lattice temperature~\cite{Butov2001}. At the same time, $\delta t \sim \tau$ and $\tau_{\rm w} << \tau$ enable the density staying high and nearly constant during the measurements. (ii) The measurements are performed $\sim 50$~$\mu$m away from the edge of the mesa-shaped laser excitation spot (Fig.~S1b in SI). This separation further facilitates cooling of the photoexcited e-h system. At the same time, the density in the signal detection region does not drop substantially in comparison to the excitation region since the separation is shorter than the I-EHP (and IX) propagation length in the heterostructure (Fig.~S1b). To further reduce the heating of e-h system, the laser excitation is resonant to the direct exciton energy ($E_{\rm ex} \sim 1.545$~eV). The resonant excitation increases absorption for a fixed $P_{\rm ex}$ and minimizes the energy of photoexcited e-h pairs~\cite{Kuznetsova2012}. The laser pulses are 800~ns on, 400~ns off. The off time is longer than $\delta t$ to enable the cooling, yet is as short as possible, just longer than $\delta t + \tau_{\rm w}$, to enhance the density for a given $P_{\rm ex}$. 

\begin{figure}
\begin{center}
\includegraphics[width=8.5cm]{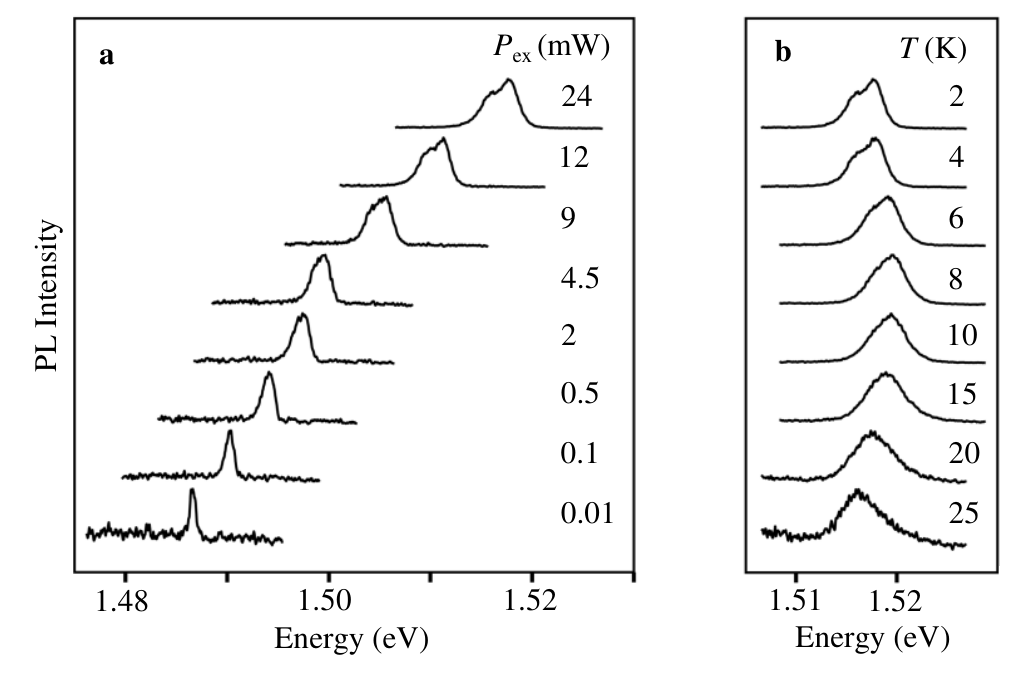}
\caption{Density and temperature dependence of PL spectra. (a) PL spectra vs. $P_{\rm ex}$. $T = 2$~K. The e-h densities estimated from the energy shift $n \sim 11, 9, 7, 5, 4, 3, 1, 0.3 \times 10^{10}$~cm$^{-2}$ (from top to bottom). The spectra show the crossover from IXs at low densities to I-EHP with the Fermi edge singularity at high densities. (b) PL spectra vs. temperature. $P_{\rm ex} = 24$~K. The Fermi edge singularity vanishes at high temperatures.
}
\end{center}
\label{fig:spectra}
\end{figure}

In the experiments, the densities of photoexcited e-h system are controlled by $P_{\rm ex}$ from the low-density IX regime to the high density I-EHP regime. In the high-density regime, we observed a broad I-EHP line with a linewidth exceeding the IX binding energy~\cite{Lozovik1997, Sivalertporn2012} and increasing with density (Figs.~1 and 2a). The simulation of the I-EHP PL line without taking into account the Fermi edge singularity due to the Cooper-pair-like excitons at the Fermi energy are presented in Fig.~S4 in SI. These simulations show step-like spectra with the linewidth $\Delta \sim E_{\rm Fe} + E_{\rm Fh}$, similar to the spectra of spatially direct EHP in single QWs in earlier studies~\cite{Butov1991}.

At high temperatures, the I-EHP PL line (Fig.~1c bottom) is typical for plasmas above the condensation temperature~\cite{Butov1991} and the lineshape is consistent with the simulations with no Fermi edge singularity (Fig.~S4). At low temperatures, we observed a strong enhancement of the PL intensity at the Fermi energy of cold plasma (Fig.~1c top) that evidences the emergence of excitonic Fermi edge singularity. The temperature and density dependence of the spectra and the measurements of the first order coherence function, presented below, are consistent with the many-body origin of this enhancement. 

At the lowest densities, the IX linewidth $\sim 0.6$~meV (Fig.~2a). The small IX linewidth indicates a low disorder in the heterostructure. With increasing e‐h density, we observe a transition from the ultracold gas of IXs, hydrogen‐like pairs of separated electrons and holes, with the narrow PL line at low e‐h densities to the ultracold I-EHP with the Fermi edge singularity at high e‐h densities (Fig.~2a). The transition is smooth, consistent with the theory predicting a smooth transition from BEC to BCS-like exciton condensate with increasing density~\cite{Comte1982}. 

The overall shift of the PL energy (Fig.~2a) is caused by the separation between the electron and hole layers and can be approximated by the "capacitor" formula $\delta E = 4\pi e^2 d n / \varepsilon$, where $d$ is the separation between the layers, $\varepsilon$ the dielectric constant~\cite{Yoshioka1990}. This approximation becomes increasingly more accurate with increasing density~\cite{Choksy2021}. The e-h density $n$ estimated from the energy shift $\delta E$ is close to $n$ estimated from the plasma PL linewidth $\Delta \sim E_{\rm Fe} + E_{\rm Fh} = \pi \hbar^2 n (1/m_{\rm e} + 1/m_{\rm h}$), where $m_{\rm e}$ and $m_{\rm h}$ are the electron and hole effective masses (Fig.~S3 in SI).

\begin{figure}
\begin{center}
\includegraphics[width=8.5cm]{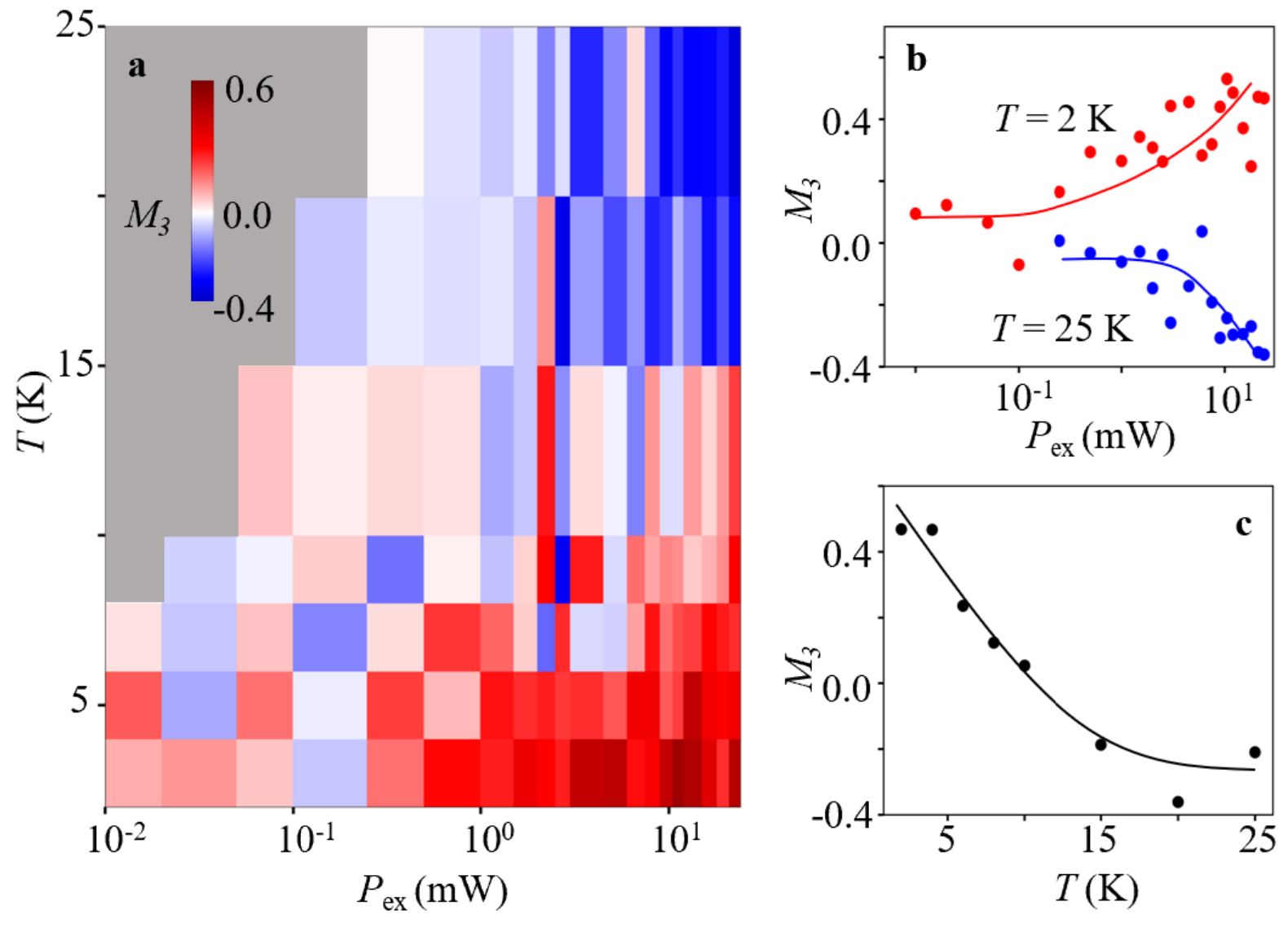}
\caption{The spectrum skewness $M_3$. (a) $M_3$ vs. $P_{\rm ex}$ and temperature. (b) $M_3$ vs. $P_{\rm ex}$ at $T = 2$~K and 25~K. (c) $M_3$ vs. temperature at $P_{\rm ex} = 24$~mW. The lines are guides to the eye. The Fermi edge singularity characterized by high positive $M_3$ is observed in dense I-EHP at low temperatures.
}
\end{center}
\label{fig:spectra}
\end{figure}

The Fermi edge singularity vanishes with increasing temperature (Fig.~2b). This is quantified in Fig.~3 by the spectrum skewness $M_3$~\cite{skewness}. A higher intensity at the high-energy (low-energy) side of the PL line, such as in the top (bottom) spectrum in Fig.~1c, corresponds to positive (negative) $M_3$. Figure 3 shows that the Fermi edge singularity characterized by high positive $M_3$ is observed in the dense I-EHP at low temperatures, that is in the high-$P_{\rm ex}$ -- low-$T$ part of the $P_{\rm ex} - T$ diagram. A similar $n - T$ diagram with the density $n$ estimated from the shift $\delta E$ is shown in Fig.~S5 in SI.

We also measured the coherence length by shift-interferometry. Similar measurements for IXs in the low-density regime detected IX spontaneous coherence and, in turn, the IX BEC in earlier studies~\cite{High2012}. In the shift-interferometry measurements, the emission images produced by each of the two arms of Mach-Zehnder interferometer are shifted with respect to each other to probe the interference between the emission of I-EHP (or IXs) spatially separated by $\delta x$ in the layer plane. The amplitude of interference fringes gives the first order coherence function $g_1(\delta x)$ and the width of $g_1(\delta x)$, the coherence length, quantifies spontaneous coherence in the system~\cite{High2012, Fogler2008}. 

The coherence vanishes with increasing temperature (Fig.~4a). With increasing $P_{\rm ex}$ and, in turn, the density, the coherence length first increases, reaches maximum, and then reduces (Fig.~4b). The density dependence is qualitatively consistent with the theory predicting that coherence increases with density in the BEC regime, reaches maximum at the BEC--BCS crossover, and reduces with density in the BCS regime~\cite{Comte1982}. 

According to the theory, the BEC--BCS crossover should occur at the densities close to the density of the Mott transition $n_{\rm M} \sim 0.2/a_{\rm B}^2$~\cite{DePalo2002, Schleede2012, Maezono2013, Fogler2014}. The density $n \sim 4 \times 10^{10}$~cm$^{-2}$ where the maximum coherence is observed (Fig.~4b) is qualitatively consistent with this theoretical estimate. For instance, for $a_{\rm B} \sim 20$~nm estimated for IXs in Ref.~\cite{Dignam1991}, $n_{\rm M} \sim 0.2/a_{\rm B}^2 \sim 5 \times 10^{10}$~cm$^{-2}$. In the density range corresponding to the onset of the BCS regime $n \gtrsim 4 \times 10^{10}$~cm$^{-2}$ ($P_{\rm ex} \gtrsim 2$~mW), the PL lineshape start revealing the intensity enhancement at the high-energy side, indicating the Fermi edge singularity (Fig.~2a). 

\begin{figure}
\begin{center}
\includegraphics[width=8.5cm]{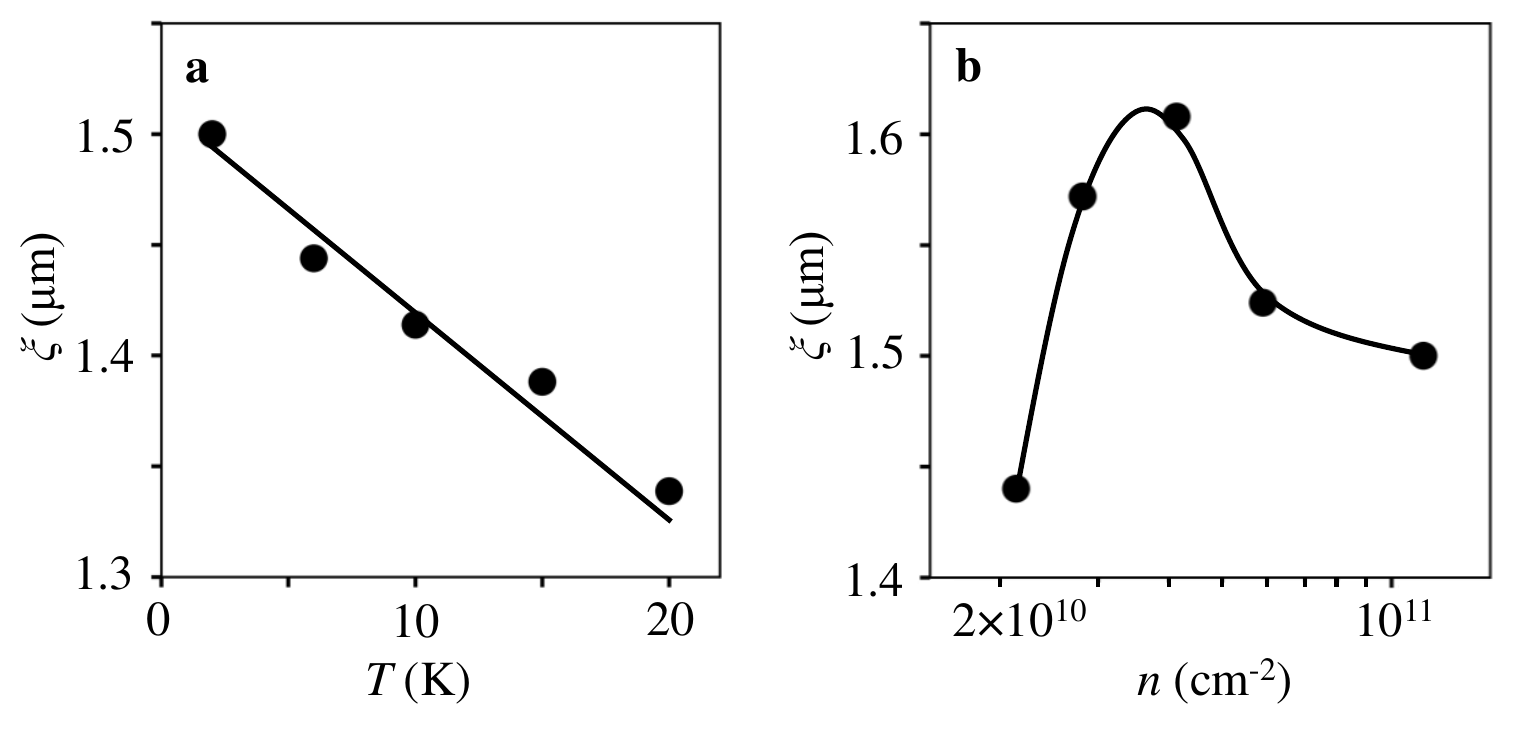}
\caption{Density and temperature dependence of the coherence length. (a) $\xi$ vs. temperature. The density $n$ estimated from the energy shift $n \sim 10^{11}$~cm$^{-2}$. $P_{\rm ex} = 24$~mW. (b) $\xi$ vs. $n$. $P_{\rm ex} = 0.2, 0.5, 2.5, 7.5, 24$~mW (from left to right). $T = 2$~K. The lines are guides to the eye.
}
\end{center}
\label{fig:spectra}
\end{figure}

The coherence length (Fig.~4) reaches significantly higher values than in a classical gas [$\xi_{\rm classical} \sim \lambda_{\rm dB} \sim 0.1$~$\mu$m at $T = 2$~K for IXs in GaAs CQW, $\lambda_{\rm dB} = (2 \pi \hbar^2/ mT)^{1/2}$ is the thermal de Broglie wavelength]. The maximum $\xi$ (Fig.~4) is smaller than in the measurements of IXs in the low-density regime in the heterostructure with smaller $d$, where $\xi$ reaches several microns~\cite{High2012}. A higher $\xi$ in that work may be related, in particular, to a weaker dipolar interaction and a specific electro-optical IX generation with holes optically generated and electrons electronically injected in localized areas~\cite{High2012}.

A relation of the studied system to other systems is outlined below. The Fermi edge singularity in neutral e-h system due to Cooper-pair-like excitons at the Fermi energy and BCS-like exciton condensation is related to excitonic insulators~\cite{Keldysh1965, DesCloizeaux1965, Kozlov1965, Jerome1967}. In contrast to optically created e-h systems in semiconductors, such as the system considered in this work, the excitonic insulators generally form in semimetals or in narrow-gap semiconductors with no optical generation. The nature of BCS-like exciton condensates in optically created e-h systems and excitonic insulators in semimetals is similar. Excitonic insulators are actively studied~\cite{Eisenstein2004, Rontani2005, Du2017, Kogar2017, Liu2017, Li2017, Burg2018, Sun2021, Ataei2021, Kim2021, Liu2022, Jia2022, Sun2022, Gu2022, Zhang2022, Chen2022}. 

The other system, which allows studying the BEC--BCS crossover, is a system of ultracold atoms with controlled interactions~\cite{Zwierlein2005}. In comparison, in the ultracold e-h system studied here, the density and, in turn, the parameter $n a_{\rm B}^2$ is controlled. The density increase allows to go from the low-density BEC regime to the high-density BCS regime and the regimes are revealed by the distinct PL lineshapes with the high-density BCS regime characterized by the Fermi edge singularity due to the Cooper-pair-like excitons at the Fermi energy. 

In summary, we found a strong enhancement of photoluminescence intensity at the Fermi energy of the neutral dense ultracold e-h system that evidences the emergence of excitonic Fermi edge singularity. 

\section{Acknowledgments}
We thank Michael Fogler, Lewis Fowler-Gerace, Jason Leonard, and Lu Sham for discussions. The spectroscopy studies were supported by NSF Grant No. 1905478. The coherence studies were supported by DOE Office of Basic Energy Sciences under Award No. DE-FG02-07ER46449. The heterostructure growth was funded by the Gordon and Betty Moore Foundation’s EPiQS Initiative, Grant GBMF9615 to L.N.~Pfeiffer, and by the National Science Foundation MRSEC grant DMR~2011750 to Princeton University.

\vskip 5 mm
$^*$equal contribution

\end{document}


\date{\today}

\title{Supporting Information for\\
Fermi edge singularity in neutral electron-hole system}

\author{D.~J.~Choksy$^*$}
\affiliation{Department of Physics, University of California at San Diego, La Jolla, CA 92093, USA}
\author{E.~A.~Szwed$^*$}
\affiliation{Department of Physics, University of California at San Diego, La Jolla, CA 92093, USA}
\author{L.~V.~Butov} 
\affiliation{Department of Physics, University of California at San Diego, La Jolla, CA 92093, USA}
\author{K.~W.~Baldwin}
\affiliation{Princeton University, Princeton, New Jersey 08544, USA}
\author{L.~N.~Pfeiffer}
\affiliation{Princeton University, Princeton, New Jersey 08544, USA}

\begin{abstract}
\noindent
\end{abstract}

\maketitle
\renewcommand*{\thefigure}{S\arabic{figure}}

\subsection{CQW heterostructure}

The CQW heterostructure (Fig.~1b) is grown by molecular beam epitaxy. CQW consists of two 15-nm GaAs QWs separated by a 4-nm Al$_{\rm 0.33}$Ga$_{\rm 0.67}$As barrier. $n^+$ GaAs layer with $n_{\rm Si} \sim 10^{18}$~cm$^{-3}$ serves as a bottom electrode. The CQW is positioned 100~nm above the $n^+$ GaAs layer within the undoped 1-$\mu$m-thick Al$_{\rm 0.33}$Ga$_{\rm 0.67}$As layer. The CQW is positioned closer to the homogeneous bottom electrode to suppress the fringing in-plane electric field in excitonic devices~\cite{Hammack2006}. The top semitransparent electrode is fabricated by applying 2-nm Ti and 7-nm Pt on a 7.5-nm GaAs cap layer. Applied gate voltage $V_{\rm g} = -2.5$~V creates an electric field in the $z$ direction. 

\begin{figure}
\begin{center}
\includegraphics[width=6.5cm]{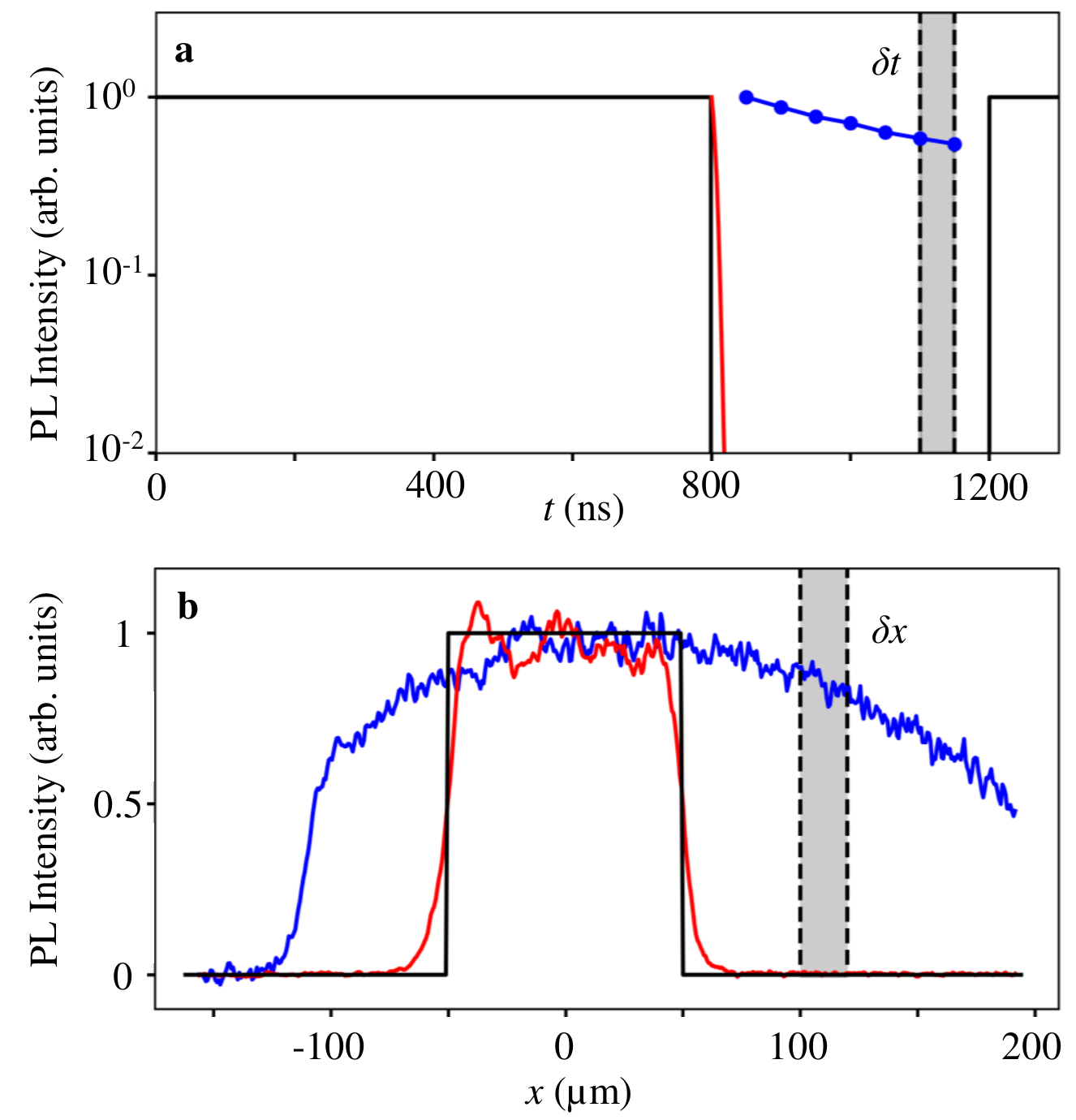}
\caption{Optical measurements. (a) The e-h system is generated by laser pulses 800~ns on, 400~ns off (shown schematically by blue line). The measurements are performed $\delta t = 300$~ns after the laser excitation pulse within $\tau_{\rm w} = 50$~ns window (gray area). (b) The laser excitation spot is mesa-shaped (shown schematically by blue line). The measurements are performed $\sim 50$~$\mu$m away from the edge of the mesa-shaped laser excitation spot within $\sim 20$~$\mu$m window (gray area). $x \sim - 100$~$\mu$m corresponds to the device edge. The DX PL (red line) closely follows the laser excitation in time (a) and space (b) due to the short DX lifetime. The I-EHP PL (black line and dots) extends in time (a) and space (b) due to the long I-EHP lifetime. 
}
\end{center}
\label{fig:spectra}
\end{figure}

\begin{figure}
\begin{center}
\includegraphics[width=6.5cm]{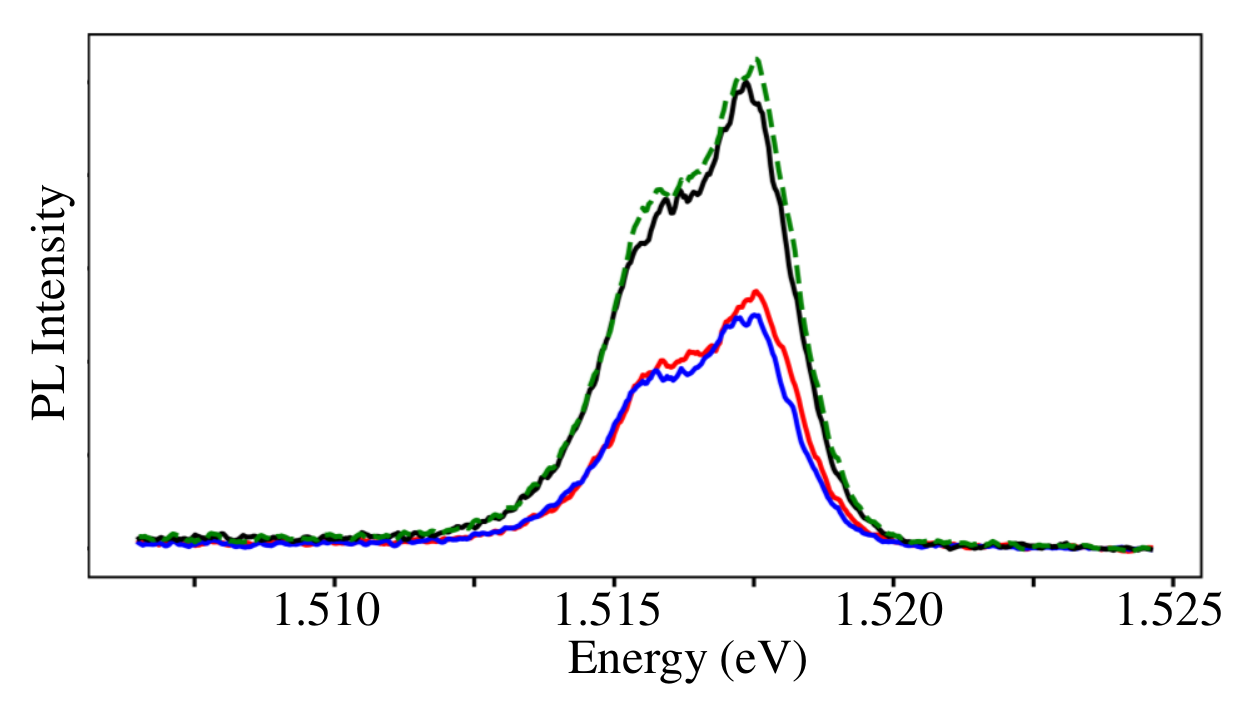}
\caption{Variation of the spectra within the signal accumulation window. The I-EHP spectrum measured during the 50~ns window (black line) and during the first half (red line) and within the second half (blue line) of the window. The sum of the spectra within the half-windows (green dashed line) is close to the spectrum within the window (black line). 
These measurements show that the spectrum variation during the window is negligibly small.
}
\end{center}
\label{fig:spectra}
\end{figure}

\subsection{Optical measurements}

The experiments are designed to facilitate lowering the temperature of the optically generated e-h system, as outlined in the main text. The e-h system is generated by a Ti:Sapphire laser resonant to the direct exciton (DX) energy ($E_{\rm ex} \sim 1.545$~eV). An AOM is used for making laser pulses 800~ns on, 400~ns off (Fig.~S1a). The measurements are performed $\delta t = 300$~ns after the laser excitation pulse within $\tau_{\rm w} = 50$~ns window (Fig.~S1a). The mesa-shaped laser excitation spot with $\sim 100$~$\mu$m diameter is formed using an axicon. The measurements are performed $\sim 50$~$\mu$m away from the edge of the mesa-shaped laser excitation spot within $\sim 20$~$\mu$m window (Fig.~S1b). The DX PL closely follows the laser excitation in time (Fig.~S1a) and space (Fig.~S1b) due to the short DX lifetime. The I-EHP (and IX) PL extends in time (Fig.~S1a) and space (Fig.~S1b) due to the long I-EHP (and IX) lifetime. 

The 50~ns window is long enough to collect sufficient I-EHP (or IX) signal yet much shorter than the I-EHP (or IX) lifetime so the signal variation during the window is negligibly small. To verify this, the measurements were performed within the first half and within the second half of the window and these measurements show similar spectra~(Fig.~S2). 

The PL spectra are measured using a spectrometer with resolution 0.2~meV and a liquid-nitrogen-cooled CCD coupled to a PicoStar HR TauTec time-gated intensifier. The experiments are performed in a variable-temperature 4He cryostat.

\subsection{PL energy shift and linewidth}

\begin{figure}
\begin{center}
\includegraphics[width=7.5cm]{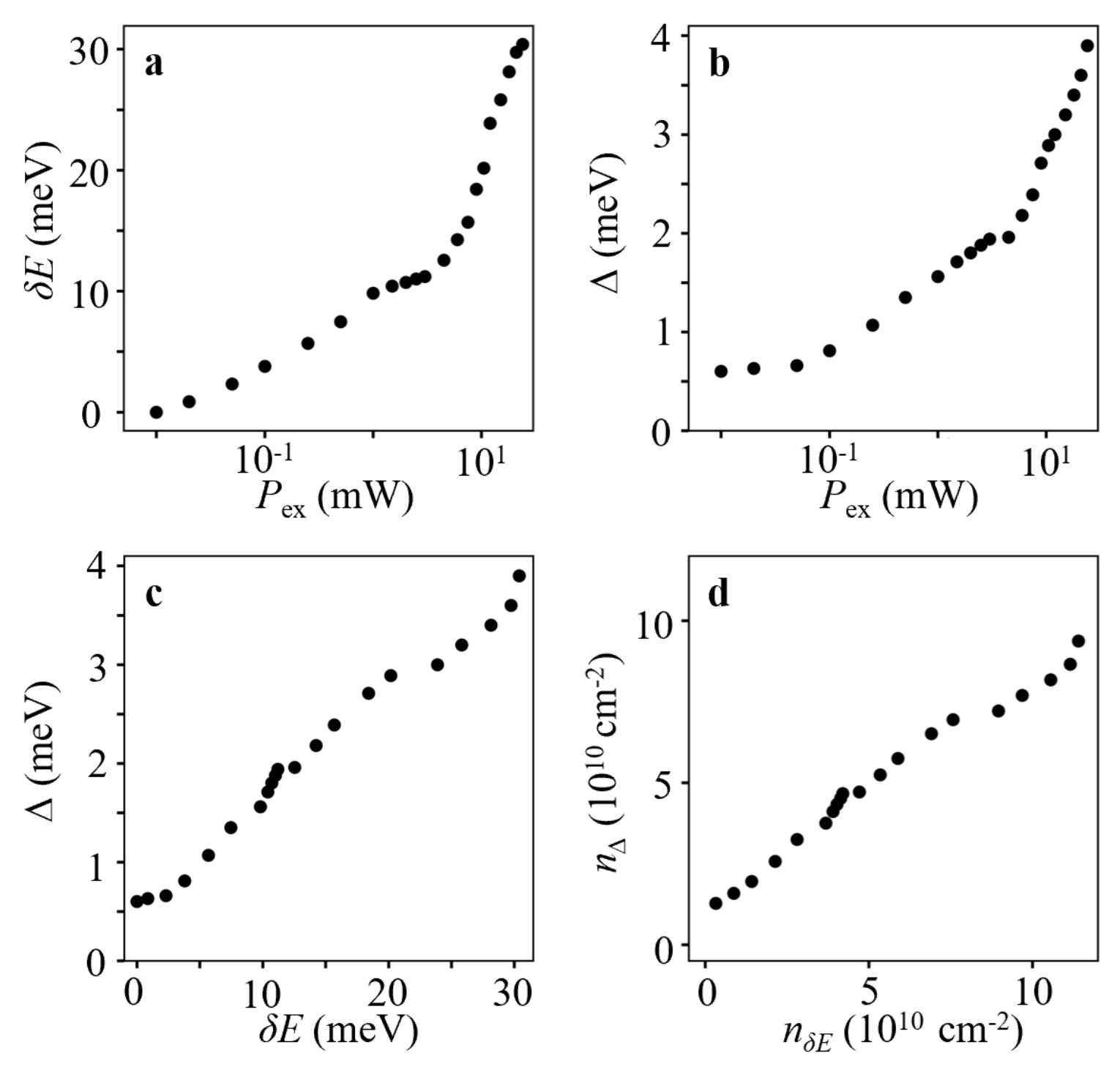}
\caption{PL energy shift and linewidth. 
(a) The PL energy shift $\delta E$ vs. $P_{\rm ex}$. $\delta E$ is counted from the IX energy at the lowest $P_{\rm ex}$. 
(b) The PL linewidth $\Delta$ (full-width-half-maximum) vs. $P_{\rm ex}$. 
(c) $\Delta$ vs. $\delta E$. 
(d) $n$ estimated from $\Delta$ vs. $n$ estimated from $\delta E$. 
$T = 2$~ K for all data.
The estimates use $\delta E = 4\pi e^2 d n / \varepsilon$ and $\Delta \sim E_{\rm Fe} + E_{\rm Fh} = \pi \hbar^2 n (1/m_{\rm e} + 1/m_{\rm h})$. 
The estimates from $\delta E$ and from $\Delta$ give similar $n$.
}
\end{center}
\label{fig:spectra}
\end{figure}

\begin{figure}
\begin{center}
\includegraphics[width=5cm]{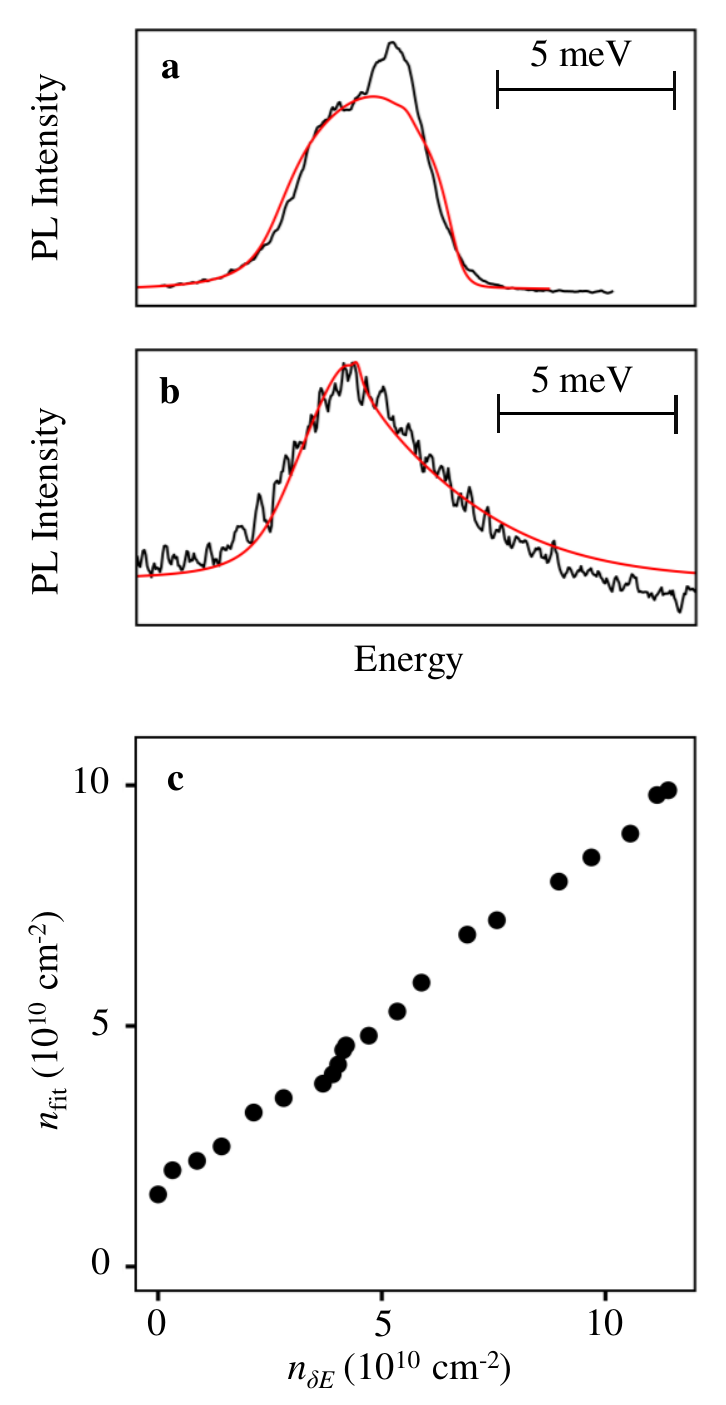}
\caption{Simulations of I-EHP PL spectra.  
(a,b) The simulated I-EHP PL spectra without taking into account the Fermi edge singularity (red lines) and the measured I-EHP PL spectra (black lines). 
$T = 2$~K and $P_{\rm ex} = 24$~mW, $T_{\rm fit} = 2$~K and $n_{\rm fit} = 10^{11}$ cm$^{-2}$ (a).
$T = 25$~K and $P_{\rm ex} = 24$~mW, $T_{\rm fit} = 25$~K and $n_{\rm fit} = 7.6 \times 10^{10}$ cm$^{-2}$ (b).
(c) $n$ estimated from the PL spectrum fit vs. $n$ estimated from $\delta E$, $T = 2$~K.
}
\end{center}
\label{fig:spectra}
\end{figure}

The density $n$ in I-EHP can be estimated from the PL energy shift $\delta E$ using the "capacitor" formula $\delta E = 4\pi e^2 d n / \varepsilon$. $n$ in I-EHP can be also estimated from the PL linewidth $\Delta \sim E_{\rm Fe} + E_{\rm Fh} = \pi \hbar^2 n (1/m_e + 1/m_h)$. The estimates from $\delta E$ and from $\Delta$ give similar $n$ in the high-density regime, $n \gtrsim 4 \times 10^{10}$~cm$^{-2}$ (Fig.~S3d). The estimates extended to the low-density regime show a deviation from this similarity, increasing for lower densities (Fig.~S3d). The deviation is expected since in the low-density regime, the equation for $\delta E$ is less accurate and $\Delta$ is determined by the homogeneous and inhomogeneous IX broadening, as outlined in the main text.

The I-EHP PL spectra are simulated without taking into account the Fermi edge singularity and compared with the measured I-EHP PL spectra (Fig.~S4). The simulations are outlined below. Due to the small photon momentum, the optical transition occur for the same absolute values of electron and hole momenta $k_{\rm e} = k_{\rm h}$. For the constant 2D density of states, the PL intensity at energy $E_i = \frac{\hbar^2 k_i^2}{2m_{\rm e}} + \frac{\hbar^2 k_i^2}{2m_{\rm h}}$ is determined by the product of the electron and hole distribution functions $I(E_i) \propto f_{\rm e}(k_i)f_{\rm h}(k_i)$, where $E_i$ is counted from the lowest PL energy, $k_i = k_{\rm e} = k_{\rm h}$, $f_{\rm e,h} = \left( \exp{\frac{\hbar^2 \left(k_i^2 - k_{\rm F}^2\right)}{2m_{\rm e,h} k_{\rm B}T}} +1 \right)^{-1}$, $k_{\rm F}$ the Fermi momentum. For low temperatures $k_{\rm B}T << E_{\rm Fe}, E_{\rm Fh}$, the PL line $I(E_i)$ is step-like with the sharp steps both on the low-energy side and the high-energy side and the width $\Delta \sim E_{\rm Fe} + E_{\rm Fh} = E_{\rm F}$. The step sharpness on the high-energy side is determined by the temperature. To account for the finite step sharpness on the low-energy side the following approximation is used. $I(E_i)$ is convolved with $\frac{\gamma(E_i)}{\pi} \frac{1}{(E - E_i)^2 - \gamma(E_i)^2}$ describing the damping of one-particle states~\cite{Landsberg1966}, where the broadening parameter $\gamma(E_i)$ is assumed to decrease to zero at the Fermi level as $(E_{\rm F} -  E_i)^2$~\cite{DasSarma2021}. 

At high temperatures, the simulated and measured I-EHP spectra are close, with the intensity reduction at the high-energy side following the thermal distribution (Fig.~S4b). At low temperatures, the simulations show step-like I-EHP spectra with the linewidth $\Delta \sim E_{\rm Fe} + E_{\rm Fh}$, similar to the spectra of spatially direct EHP in single QWs~\cite{Butov1991}, and the measured I-EHP PL is strongly enhanced at the Fermi energy in comparison to the simulations due to the Fermi edge singularity (Fig.~S4a).

For the dense I-EHP, the density $n_{\rm fit}$ estimated from the PL spectrum fit is close to the density estimated from the energy shift $\delta E$~(Fig.~4c). $n_{\rm fit}$ (Fig.~4c) is close to the density estimated from the PL linewidth $\Delta$ (Fig.~3d).

\subsection{The spectrum skewness $M_3$}

Figure~S5 shows the spectrum skewness $M_3$ vs. temperature and density. This figure is similar to Fig.~3 showing $M_3$ vs. temperature and $P_{\rm ex}$. The density $n$ is estimated from the energy shift $\delta E = 4\pi e^2 d n / \varepsilon$. Figure~S5 shows that the Fermi edge singularity characterized by high positive $M_3$ is observed in the dense I-EHP at low temperatures, that is in the high-$n$ -- low-$T$ part of the $n - T$ diagram. 

\begin{figure}
\begin{center}
\includegraphics[width=8.5cm]{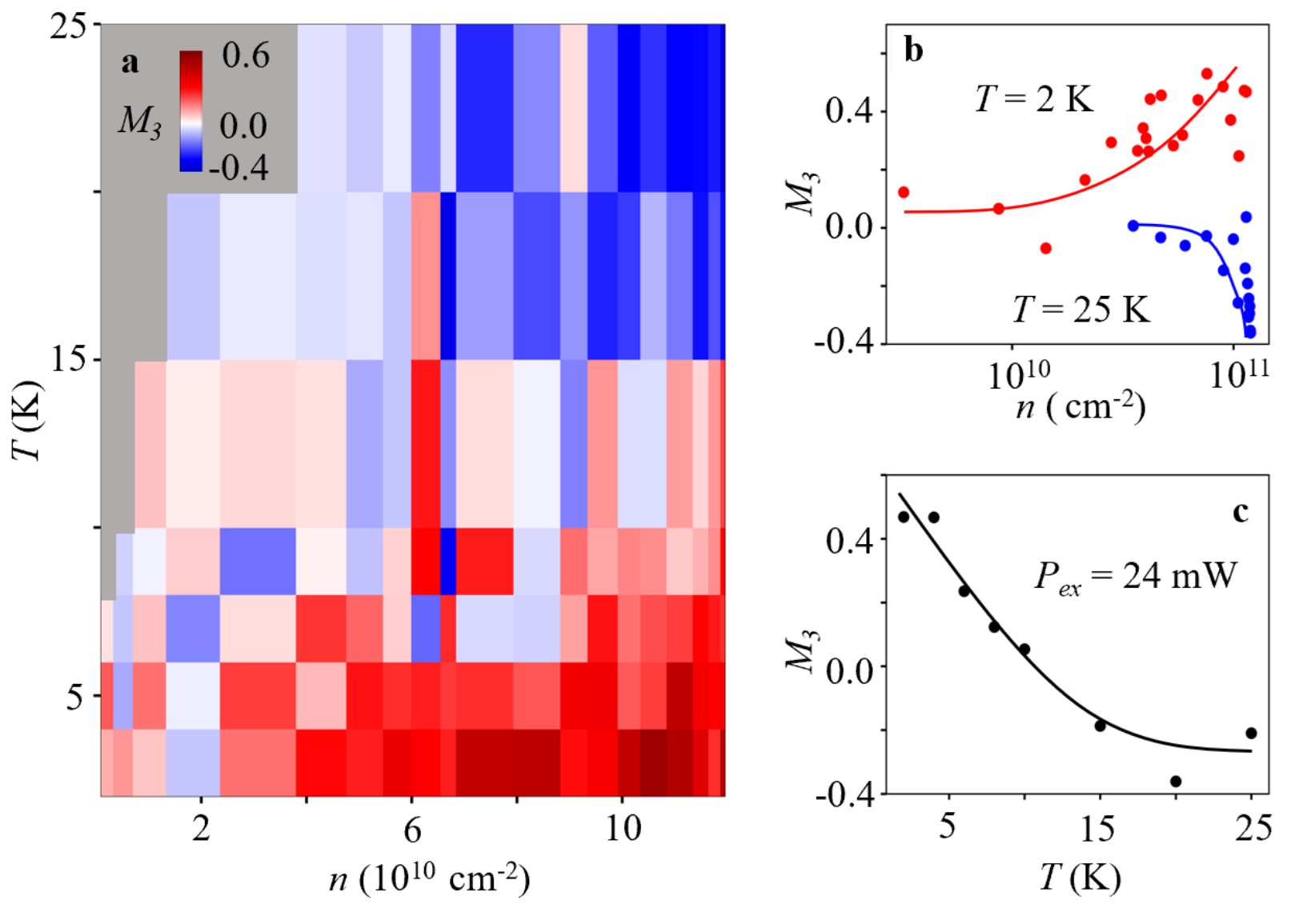}
\caption{The spectrum skewness $M_3$. (a) $M_3$ vs. density and temperature. The density $n$ is estimated from the energy shift $\delta E = 4\pi e^2 d n / \varepsilon$.
(b) $M_3$ vs. $n$ at $T = 2$~K and 25~K. (c) $M_3$ vs. temperature at $n = 10^{11}$~cm$^{-2}$. The lines are guides to the eye. The Fermi edge singularity characterized by high positive $M_3$ is observed in dense I-EHP at low temperatures.
}
\end{center}
\label{fig:spectra}
\end{figure}

\subsection{Shift-interferometry measurements}

In the shift-interferometry measurements, a Mach-Zehnder interferometer is added in the signal detection path as in Ref.~\cite{High2012}. The spectrometer grating is replaced by a mirror and an interference filter of linewidth 5~nm adjusted to the I-EHP (or IX) PL wavelength is added to select the entire I-EHP (or IX) PL line for the studied $P_{\rm ex}$ and $T$. The rest of the laser excitation and signal detection, outlined above, is kept unchanged. 

The emission images produced by each of the two arms of the Mach-Zehnder interferometer are shifted relative to each other along $x$ to measure the interference between the emission of I-EHPs (or IXs), which are separated by $\delta x$ in the layer plane. $I_{\rm interf} = (I_{12} - I_1 - I_2)/(2\sqrt{I_1 I_2})$ is calculated from the measured PL intensity $I_1$ for arm 1 open, $I_2$ for arm 2 open, and $I_{12}$ for both arms open (Fig.~S6a). In "the ideal experiment", the amplitude of interference fringes $A_{\rm interf}(\delta x)$ gives the first order coherence function $g_1(\delta x)$ and the width of $g_1(\delta x)$, the coherence length, quantifies spontaneous coherence in the system~\cite{High2012}.  In practice, the measured $A_{\rm interf}(\delta x)$ is given by the convolution of $g_1(\delta x)$ with the point-spread function (PSF) of the optical system in the experiment~\cite{Fogler2008}. The CQW contain no point source for the precise measurement of PSF. NA = 0.27 of the objective in the experiment gives a lower estimate for PSF width $\xi_{\rm PSF} \sim 0.9$~$\mu$m for the optical system. 
The dependence of $\xi$ on the parameters shows that 
the coherence length in the e-h system is 
sufficiently 
large in comparison to 
the spatial resolution of the optical system $\xi_{\rm PSF}$. 
Otherwise, the measured $\xi$ would be determined by $\xi_{\rm PSF}$ and practically would not depend on the parameters~\cite{Fogler2008}. Figure~6b shows examples of the measured $A_{\rm interf}(\delta x)$. $\xi$ are taken as the half-widths at $1/e$ height of Gaussian fits to $A_{\rm interf}(\delta x)$. 

\begin{figure}
\begin{center}
\includegraphics[width=5cm]{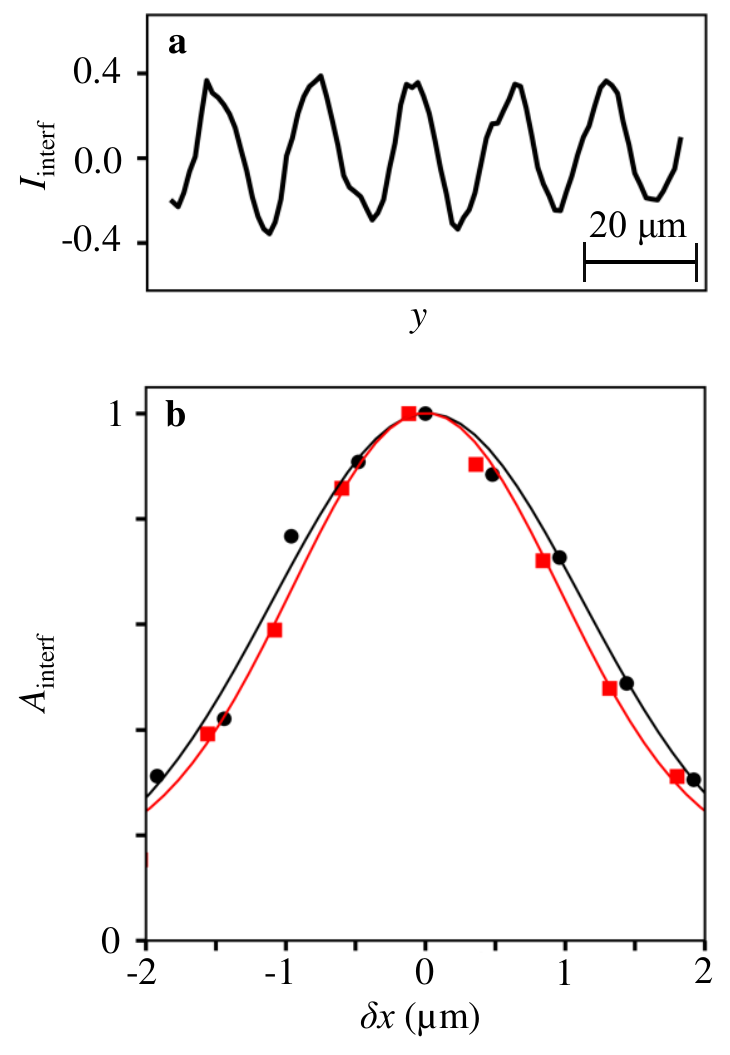}
\caption{Shift-interferometry measurements. (a) Interference fringes $I_{\rm interf}(y)$ for $\delta x = 1.5$~$\mu$m, $T = 2$~K, $P_{\rm ex} = 2.5$~mW. (b) The amplitude of interference fringes $A_{\rm interf}$ vs. $\delta x$ for $P_{\rm ex} = 2.5$~mW (black points) and 0.2~mW (red squares), $T = 2$~K. Gaussian fits are shown by the black and red lines, respectively.
}
\end{center}
\label{fig:spectra}
\end{figure}

\vskip 5 mm
$^*$equal contribution